\documentclass[journal=jpclcd,manuscript=letter]{achemso}
\usepackage{amsmath,amssymb}
\usepackage{float}

\usepackage[T1]{fontenc}
\usepackage[utf8]{luainputenc}
\usepackage{float}
\usepackage{multirow}
\usepackage{lipsum}
\usepackage{mathrsfs}
\usepackage{setspace}
\usepackage{droidsans}
\usepackage{balance}
\usepackage{times,mathptmx}
\usepackage{sectsty}
\usepackage{graphicx} 
\usepackage{lastpage}
\usepackage{adjustbox}
\usepackage{upgreek}
\usepackage{color}
\usepackage{array}
\usepackage{float}
\usepackage{fancyhdr}
\usepackage{fnpos}
\usepackage{enumerate}
\usepackage{adjustbox}
\usepackage[english]{babel}
\DeclareUnicodeCharacter{0308}{~}
\usepackage[version=3]{mhchem} 

\author{Deepika Gill}
\email{Deepika@physics.iitd.ac.in[DG]}
\author{Arunima Singh, Manjari Jain}
\author{Saswata Bhattacharya}
\email{saswata@physics.iitd.ac.in[SB]}
\phone{+91-11-2659 1359}
\affiliation[Indian Institute of Technology Delhi]
{Department of Physics, Indian Institute of Technology Delhi, New Delhi, India}
\title[An \textsf{achemso} demo]
{Exploring Exciton and Polaron Dominated Photo-physical Phenomena in Ruddlesden-Popper Phases of Ba$_{\textrm{n+1}}$Zr$_\textrm{n}$S$_{\textrm{3n+1}}$ (n=[1-3]) from Many Body Perturbation Theory}
\begin{document}

\begin{abstract}
Ruddlesden-Popper (RP) phases of Ba$_{\textrm{n+1}}$Zr$_{\textrm{n}}$S$_{\textrm{3n+1}}$ (n=[1-3]) are evolved as new promising class of chalcogenide perovskites in the field of optoelectronics, especially in solar cells. However, detailed studies regarding its optical, excitonic, polaronic and transport properties are hitherto unknown. Here, we have explored the excitonic and polaronic effect in RP phases of Ba$_{\textrm{n+1}}$Zr$_{\textrm{n}}$S$_{\textrm{3n+1}}$ (n=[1-3]) using several first-principles based state-of-the-art methodologies under the framework of Many Body Perturbation Theory. Unlike it's bulk counterpart, the optical and excitonic anisotropy are observed in Ba$_{\textrm{n+1}}$Zr$_{\textrm{n}}$S$_{\textrm{3n+1}}$ (n=[1-3]) RP phases. From Wannier-Mott approach, we show that in the RP phases of this class of chalcogenide perovskites, capturing the ionic contribution to the dielectric constant is important. We report significant ionic contribution and relatively smaller electron-phonon coupling constant for Ba$_{\textrm{n+1}}$Zr$_\textrm{n}$S$_{\textrm{3n+1}}$ in comparison to the bulk BaZrS$_3$. The exciton binding energy is found to be dependent on the presence of large electron-phonon coupling. The charge carrier mobility is maximum in Ba$_2$ZrS$_4$, computed employing deformation potential of the same. As per our analysis, the optical phonon modes are observed to dominate the acoustic phonon modes, leading to decrease in polaron mobility on increasing n in Ba$_{\textrm{n+1}}$Zr$_{\textrm{n}}$S$_{\textrm{3n+1}}$ (n=[1-3]). 
  \begin{tocentry}
  \begin{figure}[H]%
  	\includegraphics[width=1.0\columnwidth,clip]{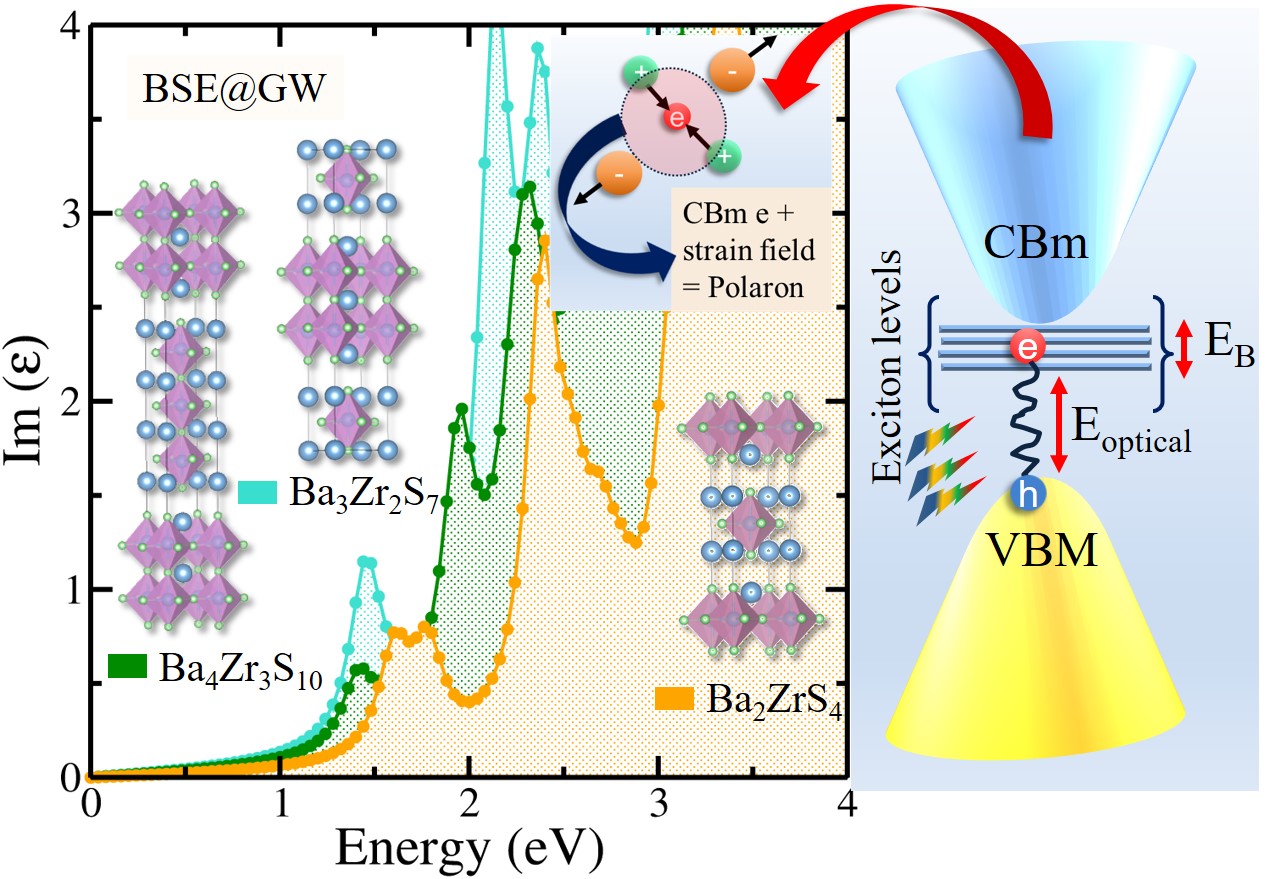}
  \end{figure}	
  \end{tocentry}
\end{abstract}
Perovskites with general chemical formula ABX$_3$ have found great attention in dielectric,  optoelectronics, and solar cell applications due to their superb ferroelectric,  piezoelectric,  superconductive and photovoltaic properties~\cite{jaffe1971piezoelectric, C6TC04830G, zhang2015advantages}.  During the last decade,  in the field of solar cell applications,  hybrid lead-halide perovskites namely,  CH$_3$NH$_3$PbX$_3$ and CH(NH$_2$)$_2$PbX$_3$ (X = Cl, Br and I) have achieved great success owing to their small band gap,  long carrier mobility,  low manufacturing cost and high power conversion efficiency~\cite{deschler2014high, shi2015low,  C4EE02988G, bi2016charge, C5CC08643D, C7TA00434F, C8CC01982G}.  However, due to the presence of organic molecules, the stability of perovskites is affected towards heat, light and moisture, thereby degrading their efficiency with time in the practical world~\cite{park2019intrinsic}.  Moreover,  presence of toxic lead in these materials makes them hazardous for the environment~\cite{babayigit2016toxicity}.  These shortcomings have hindered their practical applications. 
 
In search of alternative perovskites that can alleviate the limitations of lead-halide perovskites,  chalcogenide perovskites with S or Se-anion have been proposed for photovoltaic applications~\cite{sun2015chalcogenide}.  Several prototypical chalcogenide perovskites (viz. SrHfS$_3$~\cite{hanzawa2019material},  AZrS$_3$ (A = Sr, Ca, Ba)~\cite{lee2005synthesis, majumdar2020emerging, perera2016chalcogenide},   along with their related phases) have been synthesized successfully.  Amongst them, BaZrS$_3$ consists of earth-abundant elements and is having moderate band gap ($\sim$1.82 eV~\cite{li2019band}) ideal for photovoltaics. Moreover, it also exhibits ambipolar doping~\cite{meng2016alloying} and is stable against different environmental conditions~\cite{ravi2021colloidal}. In order to optimize the solar cell absorption,  doping at Ba/Zr-sites have been attempted in this material~\cite{wei2020ti, perera2016chalcogenide, kuhar2017sulfide}.  However, such doped/alloyed configurations seem to lack stability~\cite{sun2018chalcogenide}. Thin films of BaZrS$_3$ are also reported, which has directed the research towards its new phases named as Ruddlesden-Popper (RP) phases~\cite{comparotto2020chalcogenide}. Tremendous efforts have been invested to tune the electrical and optical properties~\cite{li2019band}.  The RP phases as an imitative of the perovskite structure are evolving as a semiconductor for optoelectronic applications~\cite{song2020structure, ghosh2020charge}.  
Their general formula is A$_{\textrm{n+1}}$B$_{\textrm{n}}$X$_{\textrm{3n+1}}$,  where perovskite structure blocks of unit cell thickness ``n'' are separated by rock salt layer AX along [001] direction.  Alternate perovskite blocks are displaced in in-plane direction by half of the unit cell.  The RP phases are included in the broad category of ``2D perovskites'' owing to their layered structural arrangement (see Fig.~\ref{1}). 

 Note that several studies assigning to the layered perovskites as ``2D perovskites'' exist in the literature, where periodic stacking of perovskite layers result in a bulk structure. Their material properties can be tuned either by substitution or dimensional reduction~\cite{gill2020understanding, gill2021high, qiu20192d}.  Due to quantum confinement effects~\cite{stanford2018emerging}, considerable change in bulk physical properties (such as bulk modulus, elastic modulus, charge carrier properties and optical properties) can be seen on reducing the dimension of material~\cite{lee2012quantum, kohn1965self}.  Research in this field is highly evolving~\cite{ghosh2020charge, tian2020two,yu2019theoretical,hohenberg1964inhomogeneous}.

 In optoelectronic materials exciton formation greatly influence the charge separation properties and hence,  excitonic parameter such as exciton binding energy (E$_\textrm{B}$) acts as an  important descriptor for optoelectronic applications.  Solar cell performance depends upon the fraction of thermally dissociated excitons into electrons and holes, giving rise to the free-charge carriers. In addition, the concept of polarons has been used to explain multiple photo-physical phenomena in these materials~\cite{park2018excited}. Polaronic effects have been suggested to play an important role in the excitation dynamics and carrier transport. The separation of free charge is also influenced by the carrier mobility. Hence understanding the effect of electron-phonon coupling in terms of polaron mobility is important. A systematic study on the excitonic and polaronic effect in the RP phases of BaZrS$_3$ is hitherto unknown. The present Letter, therefore, explores the excitonic properties along with polaronic effect in RP phases of Ba$_{\textrm{n+1}}$Zr$_{\textrm{n}}$S$_{\textrm{3n+1}}$ (n=[1-3]) under the framework of Many Body Perturbation Theory. The electron-phonon coupling is also taken care of using Frohlich model to compute the polaron mobility.

The exciton binding energy is defined as the energy required to decouple the exciton into individual electron and hole pair. Theoretically,  the exciton binding energy (E$_\textrm{B}$) is calculated by taking the difference of the energy of bounded electron-hole (e-h) pair (i.e., BSE gap) and unbounded e-h pair (i.e., GW gap). In order to determine the optical response of Ba$_{\textrm{n+1}}$Zr$_{\textrm{n}}$S$_{\textrm{3n+1}}$ (n=[1-3]) RP phases, we have calculated the imaginary part of dielectric function (Im ($\epsilon$)). Initially, we have benchmarked the exchange-correlation ($\epsilon_{\textrm{xc}}$) functional for our system.  As it is already known that single shot GW (G$_0$W$_0$) calculation strongly depends on the starting point, we need to validate the suitable starting point for G$_0$W$_0$ calculation. Note that spin-orbit coupling (SOC) effect is negligible in these systems (see Fig. S3). Hence, we have excluded SOC in our calculations (see Fig.  S5 in SI). The band gap of Ba$_2$ZrS$_4$, Ba$_3$Zr$_2$S$_7$ and Ba$_4$Zr$_3$S$_{10}$ are quite underestimated using PBE and the values are 0.61 eV, 0.42 eV and 0.34 eV, respectively.  On the other hand, the same with default parameters (viz. exact exchange = 25$\%$ and screening parameter 0.2 \AA$^{-1}$) of HSE06 are 1.39 eV, 1.18 eV and 1.08 eV, respectively. The HSE06 numbers are in good agreement with the experimental findings~\cite{li2019band}.  Further, the peak position, which is underestimated by PBE is improved by G$_0$W$_0$@PBE. Notably, the quasiparticle gaps computed using G$_0$W$_0$@PBE are overestimated in comparison to experimental band gap, since it does not take into account the exciton binding energy. However, the gaps are improved by employing BSE on top of G$_0$W$_0$@PBE. We find, the optical peak position of Ba$_2$ZrS$_4$, by performing G$_0$W$_0$@PBE and G$_0$W$_0$@HSE06 is  2.11 eV (see Fig. \ref{2}(a)) and 2.17 eV (see Fig. S4 in SI), respectively. Since, the results obtained from the consideration of HSE06 as the starting point of G$_0$W$_0$ deviate more from the experimental results, we have performed all the calculations using G$_0$W$_0$@PBE. Although G$_0$W$_0$@PBE gives better result than G$_0$W$_0$@HSE06, it is still far from the experimental optical band gap i.e., 1.33 eV.  
\begin{figure}[h]
	\centering
	\includegraphics[width=0.80\textwidth]{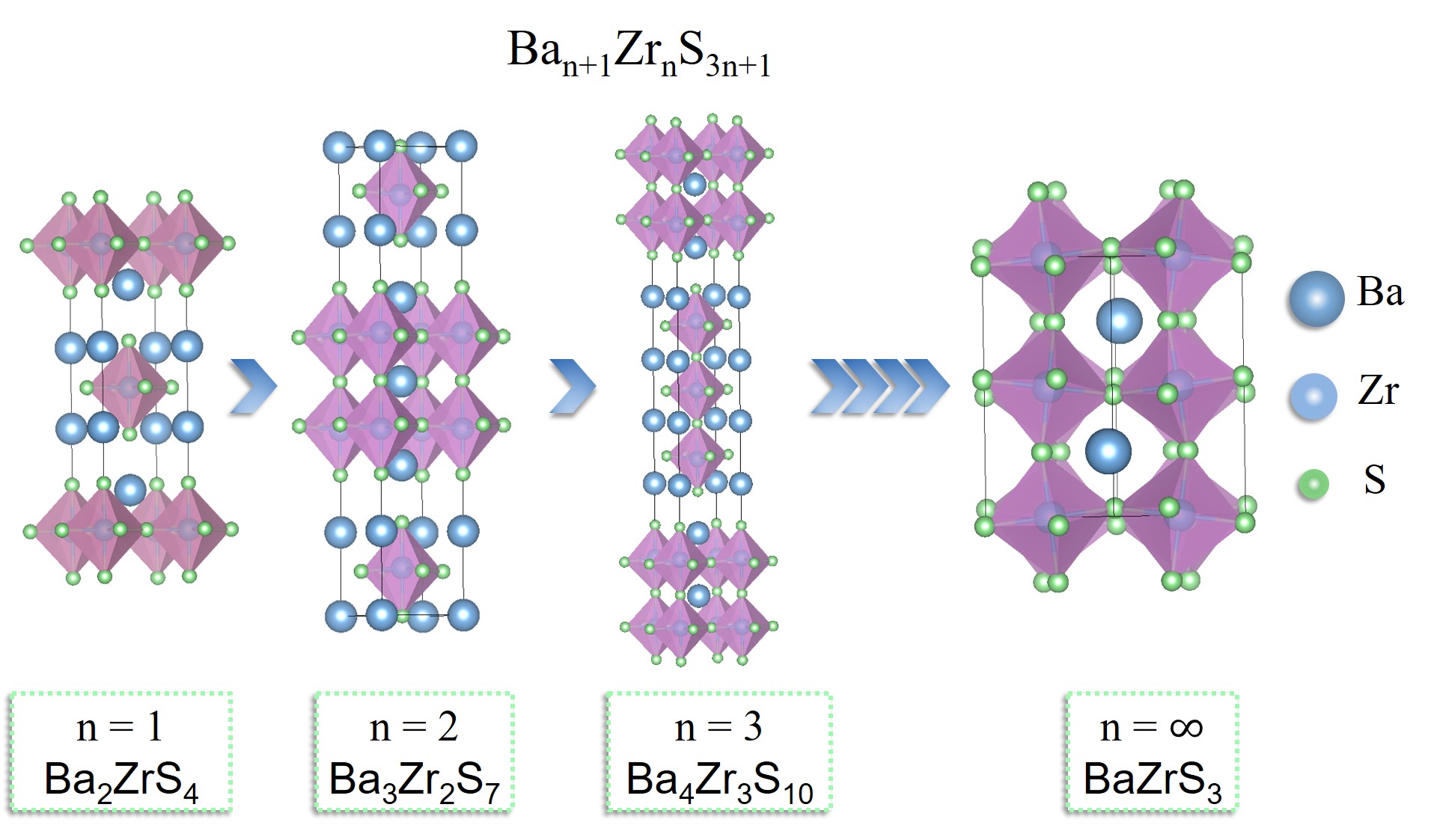}
	\caption{Optimized crystal structure of Ba$_{\textrm{n+1}}$Zr$_\textrm{n}$S$_{\textrm{3n+1}}$ (n=[1-3]) Ruddlesden-Popper phases (RP phases).}
	\label{1}
\end{figure}  

However, it gets improved when we solve the BSE to obtain optical band gap due to the fact that BSE takes into account the excitonic effect. The latter is ignored in G$_0$W$_0$ calculation to obtain accurate optical spectra. Therefore, we have performed BSE@G$_0$W$_0$@PBE to incorporate e-h interactions. Similarly, we have performed G$_0$W$_0$@PBE and BSE@G$_0$W$_0$@PBE calculations to capture the optical and excitonic effect for Ba$_3$Zr$_2$S$_7$ and Ba$_4$Zr$_3$S$_{10}$, respectively. 
We find the BSE peak position for Ba$_2$ZrS$_4$, Ba$_3$Zr$_2$S$_7$ and Ba$_4$Zr$_3$S$_{10}$ are 1.71 eV, 1.49 eV and 1.43 eV, respectively, whereas G$_0$W$_0$@PBE finds the peak at 2.11 eV, 1.82 eV and 1.69 eV, respectively (see Fig. \ref{2}(a-c)). 

It should be noted here that these numbers are highly dependent on the k-mesh and it's very challenging (even with the fastest supercomputers) to converge the BSE calculation to obtain the excitonic peak for computing E$_\textrm{B}$ with absolute accuracy. In Fig.~\ref{2}, occurrence of red-shifted peak in BSE@G$_0$W$_0$@PBE than G$_0$W$_0$@PBE signifies the excitonic effect in the considered RP phases. The computed E$_\textrm{B}$ of first bright exciton of Ba$_2$ZrS$_4$, Ba$_3$Zr$_2$S$_7$ and Ba$_4$Zr$_3$S$_{10}$ RP phases are found to be 0.40 eV, 0.33 eV and 0.26 eV, respectively. The obtained E$_\textrm{B}$ values are somewhat on the larger side. In solar cell, for easy dissociation of exciton into free charge carriers (viz.  e and h) at room temperature, a low E$_\textrm{B}$ is desirable ($k_\textrm{B}$T = 26 meV, T = 300 K). Thus the discrepancy in the BSE peak position from the experimental value~\cite{li2019band} may lead to incorrect E$_\textrm{B}$ values. Unfortunately, we have already ensured the highest possible k-mesh to compute the BSE@G$_0$W$_0$@PBE calculations. Involving a denser k-mesh is not feasible for the superstructures of RP phases for computing G$_0$W$_0$@PBE and BSE@G$_0$W$_0$@PBE -- solely due to computational limitation. Therefore, this limits us to estimate the accurate E$_\textrm{B}$ for the given systems. This is why to compute E$_\textrm{B}$, we have employed a combined state-of-the-art method comprising of Wannier-Mott and Density Functional Perturbation Theory (DFPT) approach as explained later. However, one can estimate the error bar in BSE peak position via a parameterized model for dielectric screening i.e., model-BSE approach (mBSE). This mBSE approach is computationally cheaper as compared to BSE but with similar accuracy to estimate the first peak position (see Fig. S6 in SI). Thus this enables us to sample Brillouin zone with higher number of k-mesh. As per the mBSE calculation, it is seen that a denser k-mesh sampling indeed red-shifts the BSE peak by $\sim$ 0.3 eV (see mBSE calculations details as in section V in SI).
 
Now, from the above analysis, despite having little discrepancy with the exact BSE peak position and corresponding E$_\textrm{B}$ values, it's certain that the first two excitons are bright excitons in the considered RP phases and several dark excitons also exist below the second bright exciton of these systems. From the above studies, it's also expected to have the correct trend i.e., the order of E$_\textrm{B}$ for Ba$_{\textrm{n+1}}$Zr$_\textrm{n}$S$_{\textrm{3n+1}}$ is n=1 $>$ n=2 $>$ n=3. From Fig. \ref{2}(a), we have observed the double peak character for excitonic peak of Ba$_2$ZrS$_4$ that may occur due to two nearby transitions. 
Moreover, only Ba$_2$ZrS$_4$ exhibits direct band gap, whereas for other two higher RP phases in the series band gap becomes indirect in nature. In Fig.~\ref{4},  we have shown the broadening of excitonic peak. It is well known that lifetime ($\tau$) of exciton is inversely proportional to the same and therefore,  the qualitative trend of exciton lifetime for Ba$_{\textrm{n+1}}$Zr$_\textrm{n}$S$_{\textrm{3n+1}}$ (n=[1-3]) is $\tau_{n=3}$ $>$ $\tau_{n=2}$ $>$ $\tau_{n=1}$.  Note that, here broadening is computed from mBSE approach showing contribution due to the electron-hole interaction and it does not include the electron-phonon coupling effect.  
\begin{figure}[htp]
	\centering
	\includegraphics[width=0.80\textwidth]{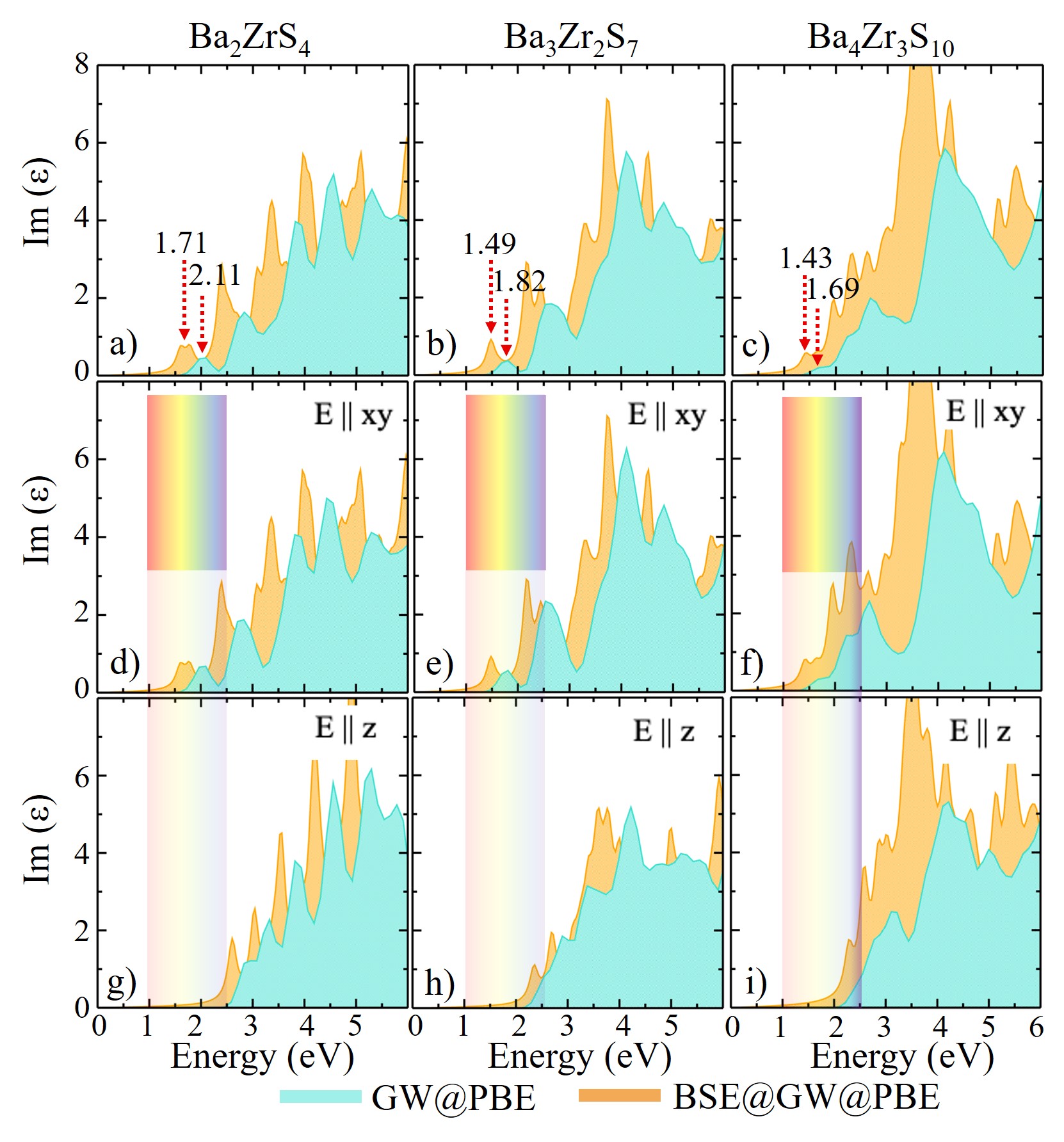}
	\caption{Imaginary part (Im ($\epsilon$)) of the dielectric function for (a) Ba$_2$ZrS$_4$, (b) Ba$_3$Zr$_2$S$_7$, (c) Ba$_4$Zr$_3$S$_{10}$ using single shot GW (G$_0$W$_0$) and BSE. Imaginary part (Im($\varepsilon$)) of dielectric function for (d) Ba$_2$ZrS$_4$ (e) Ba$_3$Zr$_2$S$_7$ and (f) Ba$_4$Zr$_3$S$_{10}$ along E $||$ xy direction and  imaginary part (Im($\varepsilon$)) of dielectric function for (g) Ba$_2$ZrS$_4$ (h) Ba$_3$Zr$_2$S$_7$ and (i) Ba$_4$Zr$_3$S$_{10}$ along E $||$ z direction, using G$_0$W$_0$ and BSE.}
	\label{2}
\end{figure}

\begin{figure}[h]
	\centering
	\includegraphics[width=0.50\textwidth]{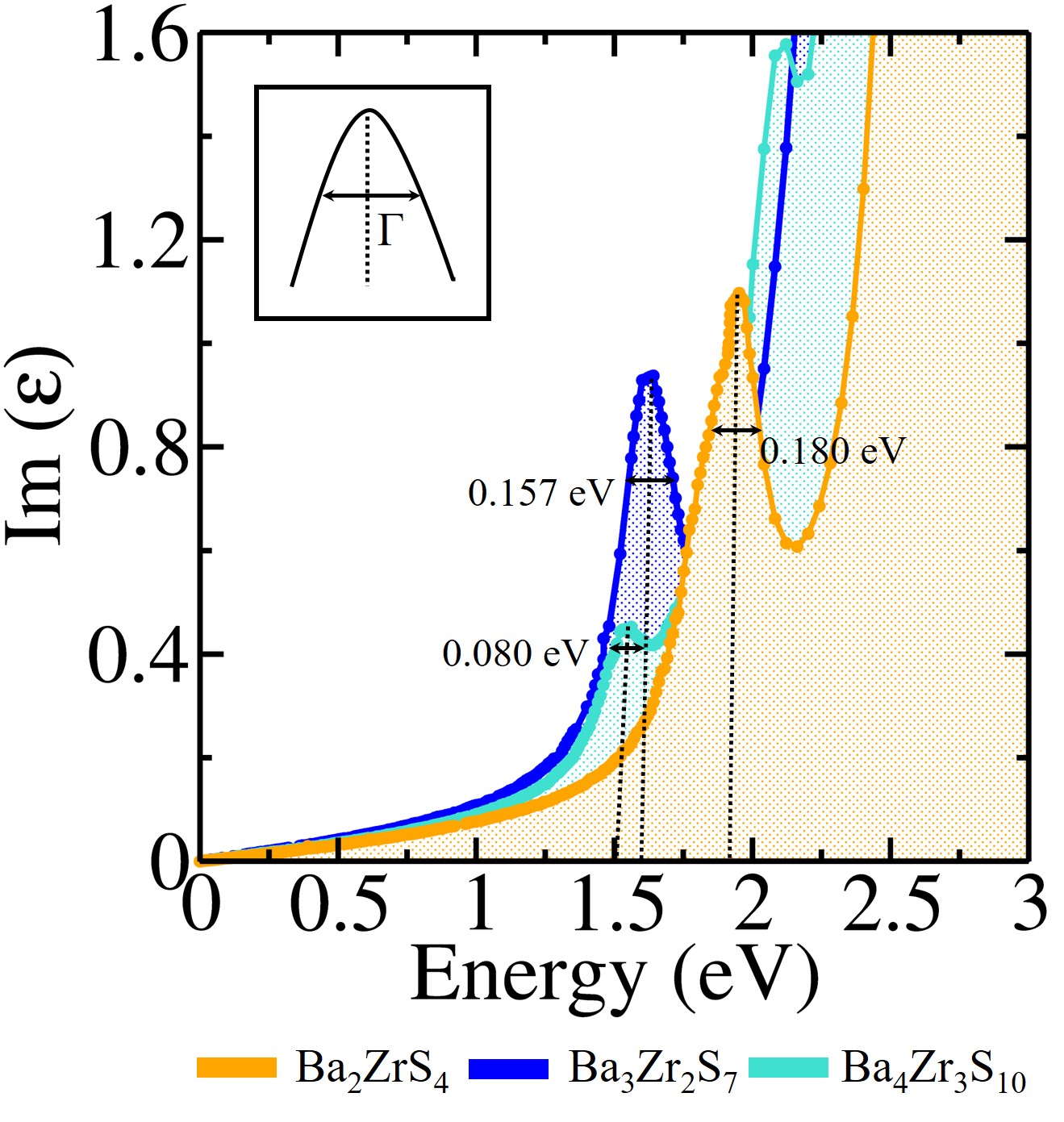}
	\caption{Full width at half maximum (FWHM) of exciton peak using mBSE approach with dense k-mesh. Broadening of exciton peak is mainly due to electron-hole interaction.}
	\label{4}	
\end{figure}

 Ba$_{\textrm{n+1}}$Zr$_\textrm{n}$S$_{\textrm{3n+1}}$ (n=[1-3]) RP phases have tetragonal structure and exhibit optical anisotropy. Hence, it is required to study their optical and excitonic properties along E$\vert\vert$xy (i.e., in-plane along x- or y-direction) and E$\vert\vert$z (i.e., out-of-plane along z-direction) direction. We have observed anisotropy in Ba$_2$ZrS$_4$, Ba$_3$Zr$_2$S$_7$ and Ba$_4$Zr$_3$S$_{10}$ which can greatly affect their performance in practical application. Therefore, it is of paramount importance to understand the anisotropic effect in the optical and excitonic properties of Ba$_{\textrm{n+1}}$Zr$_{\textrm{n}}$S$_{\textrm{3n+1}}$ (n=[1-3]) RP phases.  In Fig. \ref{2} (d-i),  we have shown optical and excitonic contribution of Ba$_{\textrm{n+1}}$Zr$_\textrm{n}$S$_{\textrm{3n+1}}$ (n=[1-3]) RP phases along different directions viz.  x, y and z.  
Employing Shockley-Queisser (SQ) criterion~\cite{shockley1961detailed} for the solar cell and other optoelectronic devices, we can remark that Ba$_{\textrm{n+1}}$Zr$_\textrm{n}$S$_{\textrm{3n+1}}$ (n=[1-3]) RP phases are optically active in in-plane (i.e., along x- and y-direction) and optically inactive in out-of-plane (i.e., z-direction).  These systems possess similar optical as well as excitonic properties along x- and y-direction (see Fig. \ref{2}(d-f)). However, along z-direction, their optical and excitonic spectra are not only blue-shifted but also the feature of G$_0$W$_0$ and BSE peaks are quite different than that in case of in-plane direction (see Fig. \ref{2}(g-i)).  It is well known that exciton lifetime is inversely proportional to the width of the exciton peak.  Hence, change in the feature of exciton peak greatly influences the excitonic parameters as well in different directions. The d-orbital contribution in valence band maximum (VBM) and conduction band minimum (CBm) could be responsible for the optical anisotropy in these systems.  Here, Zr 4d$_{xy}$  orbital is contributing maximum in the CBm (see Fig. S7 in SI), whereas in bulk BaZrS$_3$ the CBm is contributed by d$_{yz}$ and d$_{xz}$. This is why in the latter, no significant anisotropy is observed~\cite{manish-chalco}.

We have used Wannier-Mott model~\cite{waters2020semiclassical} for a simple screened Coulomb potential.  According to this model,  the E$_\textrm{B}$ for screened interacting electron-hole (e-h) pair is given by:
\begin{equation}
\begin{split}
\textrm{E}_\textrm{B}=\left(\frac{\mu}{\epsilon_\textrm{eff}^2}\right)\textrm{R}_\infty
\label{eq1}
\end{split}
\end{equation}

\noindent where, $\mu$ is the reduced mass in term of rest mass of electron, $\epsilon_\textrm{eff}$ is the effective dielectric constant (which includes electronic as well as ionic contribution to dielectric constant) and R$_\infty$ is the Rydberg constant.  The reduced mass of Ba$_2$ZrS$_4$, Ba$_3$Z$_2$S$_7$ and Ba$_4$Zr$_3$S$_{10}$ in term of electron rest mass are 0.32, 0.26 and 0.34, respectively (for more information regarding calculation of reduced mass and effective mass of electron and hole see section VII in SI).  However, in the above expression, $\epsilon_\textrm{eff}$ is still unknown for these systems. It is already reported that lattice relaxation can influence the exciton binding energy~\cite{freysoldt2014first}. For example, if $\omega_{\textrm{LO}}$ corresponds to longitudinal optical phonon frequency and E$_\textrm{B}$ $<<$ $\hbar\omega_{\textrm{LO}}$, one needs to consider the effect of lattice relaxation to compute $\epsilon_\textrm{eff}$. However, if E$_\textrm{B}$ $>>$ $\hbar\omega_{\textrm{LO}}$, the effect of lattice relaxation can be ignored as in such cases $\epsilon_\textrm{eff}$ $\rightarrow$ $\epsilon_e$, where $\epsilon_e$ is the static value of dielectric constant at high frequency that mainly consists of electronic contribution.

Therefore,  for $\epsilon_{\textrm{eff}}$, a value intermediate between the static electronic dielectric constant at high-frequency i.e., $\epsilon_e$ and the static ionic dielectric constant at low frequency i.e., $\epsilon_i$ should be considered. 
\begin{figure}[h]
	\centering
	\includegraphics[width=1.0\textwidth]{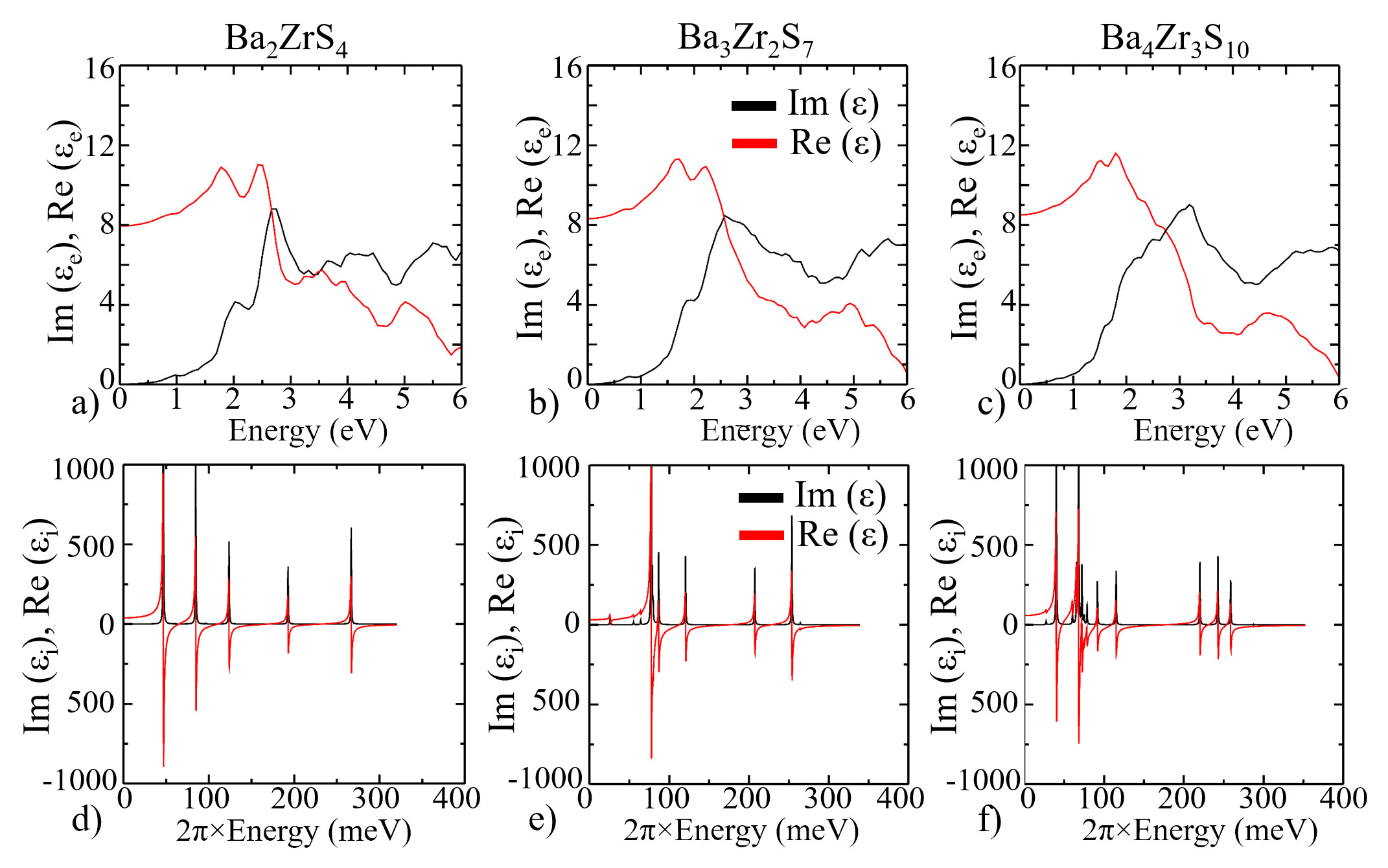}
	\caption{Electronic (Im ($\varepsilon_e$) and Re ($\varepsilon_e$)) (a)-(c) and ionic (Im ($\varepsilon_i$) and Re ($\varepsilon_i$)) (d)-(f) contribution to dielectric function for Ba$_2$ZrS$_4$,  Ba$_3$Zr$_2$S$_7$ and Ba$_4$Zr$_3$S$_{10}$. Red and black color correspond to real (Re($\varepsilon$)) and imaginary (Im($\varepsilon$)), respectively.} 
	\label{3}	
\end{figure}
In Fig. \ref{3}(a-c) and Fig. \ref{3}(d-f), we have shown the electronic and ionic contribution to the dielectric function respectively, computed using DFPT approach. The static real part of ionic dielectric constant for Ba$_2$ZrS$_4$, Ba$_3$Z$_2$S$_7$ and Ba$_4$Zr$_3$S$_{10}$ are 39.35, 31.19 and 57.59, respectively.  As per Fig. \ref{3}, a considerable increase in the static low frequency of ionic dielectric constant is attributed to the occurrence of optically active phonon modes below 10 meV.  This shows the ionic nature of the RP phases. Using electronic and ionic contribution of the dielectric constant and equation \ref{eq1}, we have calculated the upper and lower bound for the E$_\textrm{B}$ (see Table\ref{Table1}).  The effective value of the dielectric constant and hence, the binding energy lies in between these upper and lower bounds listed in Table\ref{Table1}.  
We have observed that upper bound and lower bound values of E$_\textrm{B}$ for the considered RP phases are comparable with that of APbX$_3$ (A = MA, FA; X= I, Br) perovskites~\cite{basera2020capturing, Jain2021exciton}. 
It should be noted that in bulk BaZrS$_3$ ionic contribution to dielectric constant has been observed to be negligible~\cite{manish-chalco}, whereas in RP phases significant role of ionic contribution to dielectric has been observed.  
\begin{table}[htbp]
	\caption{Electronic and ionic contribution to the dielectric constant for Ba$_{\textrm{n+1}}$Zr$_\textrm{n}$S$_{\textrm{3n+1}}$ (n=[1-3]) RP phases, where $\epsilon_e$ and $\epsilon_i$ correspond to the static value of electronic and ionic dielectric constant, respectively.  E$_{\textrm{Bu}}$ and E$_{\textrm{Bl}}$ correspond to upper and lower bound of exciton binding energy, respectively.} 
	\begin{center}
		\begin{tabular}[c]{|c|c|c|c|c|} \hline
			 Ba$_{\textrm{n+1}}$Zr$_\textrm{n}$S$_{\textrm{3n+1}}$ & $\epsilon_e$ & E$_{\textrm{Bu}}$ (meV) &  $\epsilon_i$  &   E$_{\textrm{Bl}}$ (meV)\\ \hline
			 Ba$_2$ZrS$_4$ & 8.94 & 56.00& 39.25 & 2.9   \\ \hline
			 Ba$_3$Zr$_2$S$_7$  & 8.32 &  50.14& 31.19 & 3.49  \\ \hline
			Ba$_4$Zr$_3$S$_{10}$ & 8.51 & 63.45 & 57.59 & 1.39  \\ \hline
		\end{tabular}
		\label{Table1}
	\end{center}
\end{table}

Recently, Ming \textit{et al.} have reported the effect of strain on the band gap and octahedron rotation for Ba$_2$ZrS$_4$.~\cite{ming2020octahedron}. According to their report, a significant change in the band gap and octahedron rotation is observed with the application of strain. However,  in our case of Ba$_2$ZrS$_4$ RP phase, monotonic change in the band gap (but not as large as in case of Ming \textit{et al.}) has been observed on applying upto $\pm7\%$ strain along b-axis and c-axis (see Fig.  S9 in SI). The effect on the band gap along b- and a-direction are symmetric. Further, the effect of strain on the band gap in out-of-plane direction is more significant than in in-plane direction. We have also noticed slight octahedral tilt under the application of strain along b-axis. In case of strain along c-axis, a small rotation of octahedron about c-axis has been observed. Here, we have calculated the elastic modulus of the RP phases that is given by
\begin{equation}
\begin{split}
\textrm{C}_{\textrm{3D}} = \left(\frac{1}{\textrm{V}}\right)\frac{\delta^2\textrm{E}}{\delta s^2}
\label{eq2}
\end{split}
\end{equation}
where V, s and E correspond to the volume of a unit cell, strain and total energy, respectively. We have observed that elastic modulus of the RP phases is larger than 2D RP phases~\cite{ming2020octahedron}. In order to compute carrier mobility, we have used deformation potential model~\cite{deformation, takagi1994universality, bruzzone2011ab, qiao2014high}. According to this model, the mobility of charge carrier is defined as: 
\begin{equation}
\begin{split}
\mu_{\textrm{DP}} = \frac{(8\pi)^{\frac{1}{2}}\hbar^4e\textrm{C}_{\textrm{3D}}}{3(m^*)^{5/2}(k_\textrm{B}\textrm{T})^{3/2}\textrm{E}_{l}^2}
\label{eq2}
\end{split}
\end{equation}
where T is the temperature, m$^*$ is the effective mass of charge carrier and \textit{e} is the elementary charge of electron.  
E$_l$ = $\Delta\textrm{V}/(\Delta\textrm{l}/\textrm{l}$) is the deformation potential (for more details regarding calculation of C$_{\textrm{3D}}$ and E$_l$ see section VIII in SI). 
In Table \ref{Table2}, we have listed the values of C$_{\textrm{3D}}$, E$_l$ and $\mu_{\textrm{DP}}$ for electron and hole of Ba$_{\textrm{n+1}}$Zr$_\textrm{n}$S$_{\textrm{3n+1}}$ (n=[1-3]) RP phases.  As elastic modulus decreases down the column (see Table \ref{Table2}), we can say that softness of the RP phases increases on increasing the value of n in Ba$_{\textrm{n+1}}$Zr$_\textrm{n}$S$_{\textrm{3n+1}}$ (n=[1-3]). Mobility of electron also decreases on increasing n in Ba$_{\textrm{n+1}}$Zr$_\textrm{n}$S$_{\textrm{3n+1}}$ (n=[1-3]).  

\begin{table}[htbp]
	\caption{Elastic modulus, deformation potential and predicted carrier mobility of Ba$_{\textrm{n+1}}$Zr$_\textrm{n}$S$_{\textrm{3n+1}}$ (n=[1-3]) RP phases.} 
	\begin{center}
		\begin{tabular}[c]{|c|c|c|c|} \hline
			  Ba$_{\textrm{n+1}}$Zr$_\textrm{n}$S$_{\textrm{3n+1}}$ & C$_{3\textrm{D}}$ (eV \AA$^{-3}$) & E$_l$ (eV) & $\mu_{\textrm{DP}}$ (cm$^2$ V$^{-1}$ s$^{-1}$)  \\ \hline
			 Ba$_2$ZrS$_4$ (e)& 1.23 & 6.36 & 2.16$\times$10$^4$   \\ \hline
			 Ba$_2$ZrS$_4$ (h)& 1.23 &  6.61 & 2.32$\times$10$^3$ \\ \hline
			 Ba$_3$Z$_2$S$_7$ (e)& 0.92 &  6.41 & 8.63$\times$10$^3$  \\ \hline
			 Ba$_3$Zr$_2$S$_7$ (h)& 0.92 &  6.39 & 4.72$\times$10$^1$ \\ \hline
			Ba$_4$Zr$_3$S$_{10}$ (e)& 0.67  & 6.72 & 4.22$\times$10$^3$  \\ \hline
			Ba$_4$Zr$_3$S$_{10}$ (h)& 0.67 & 6.73 & 8.27$\times$10$^1$   \\ \hline
		\end{tabular}
		\label{Table2}
	\end{center}
\end{table}
\begin{table*}[htbp]
	\caption{Polaron parameters for Ba$_{\textrm{n+1}}$Zr$_\textrm{n}$S$_{\textrm{3n+1}}$ (n=[1-3]) RP phases. } 
  \begin{center}
  		\begin{adjustbox}{width=0.70\textwidth}
		\begin{tabular}[c]{|c|c|c|c|c|c|} \hline
			 Ba$_{\textrm{n+1}}$Zr$_{\textrm{n}}$S$_{\textrm{3n+1}}$ & 1/$\epsilon^*$ & $\alpha$ &  $\textrm{m}_\textrm{P}$/$\textrm{m}^*$ &  l$_{\textrm{P}}$ (\AA) & $\mu_{\textrm{P}}$ (cm$^2$V$^{-1}$s$^{-1}$)\\ \hline
			 Ba$_2$ZrS$_4$  & 0.09 & 1.84& 1.39 & 354.30 & 164.75 \\ \hline
			 Ba$_3$Z$_2$S$_7$   & 0.08 &  2.36 & 1.53 & 307.12& 117.20   \\ \hline
			Ba$_4$Zr$_3$S$_{10}$  &0.10 & 1.77 & 1.37 &  292.49  & 76.39 \\ \hline
		\end{tabular}
	     \end{adjustbox}
		\label{Table3}
  \end{center}
\end{table*}

After analyzing the specific free volume (for details see section IX in SI), we find that study of electron-phonon coupling is important in these materials.  Also, the presence of polarization in the RP phases lays emphasis on the polaron study.
We have examined the electron-phonon interaction in our system by the mesoscopic model, viz. Frohlich's model~\cite{frohlich1954electrons, feynman1955slow, hellwarth1999mobility} for the polarons.  The dressed ``quasiparticles'', formed due to screened interaction of electron and hole by the lattice, are known as polarons.  Frohlich introduced a parameter to describe theoretically the momentum of electron in the field of polar lattice vibration. This parameter is known as dimensionless Frohlich coupling constant~\cite{biaggio1997band, frohlich1954electrons, feynman1955slow}.
 
\begin{equation}
\begin{split}
\alpha = \frac{1}{\epsilon^*}\sqrt{\frac{\textrm{R}_\textrm{y}}{\textrm{ch}\omega_{\textrm{LO}}}}\sqrt{\frac{\textrm{m}^*}{\textrm{m}_\textrm{e}}}
\label{eq3}
\end{split}
\end{equation}
where coupling constant $\alpha$ quantifies the electron-phonon coupling, m$^*$ is the effective mass of electron, m$_\textrm{e}$ is the rest mass of the electron, h is Planck's constant, c is the speed of light, $\omega_{\textrm{LO}}$ (in [cm$^{-1}$] units) is the optical phonon frequency, $1/\epsilon^*$ is the ionic screening parameter ($1/\epsilon^*$ = $1/\epsilon_{\infty}$ - $1/\epsilon_{\textrm{static}}$ were, $\epsilon_{\textrm{static}}$ and $\epsilon_{\infty}$ are static and high frequency dielectric constant) and $\textrm{R}_\textrm{y}$ is the Rydberg energy.  We have observed that electron-phonon coupling constant of considered RP phases (Table\ref{Table3}), are smaller than that of their bulk BaZrS$_3$~\cite{manish-chalco}.  Further, using the extended form of Frohlich's polaron theory, given by Feynman,  the effective mass of polaron (m$_{\textrm{P}}$)~\cite{feynman1955slow} is defined as:
\begin{equation}
\begin{split}
\textrm{m}_{\textrm{P}} = \textrm{m}^* \left(1 + \frac{\alpha}{6} +  \frac{\alpha^2}{40} + ......\right) 
\label{eq4}
\end{split}
\end{equation}
where m$^*$ is the effective mass calculated from the band structure calculations (see section VII in SI).  The polaron radii~\cite{sendner2016optical} can be calculated as follows: 
\begin{equation}
\begin{split}
\textrm{l}_\textrm{P} = \sqrt{\frac{h}{2c\textrm{m}^*\omega_{\textrm{LO}}}}
\label{eq5}
\end{split}
\end{equation} 
Polaron mobility according to the Hellwarth polaron model is defined as follows: 
 \begin{equation}
\begin{split}
\mu_{\textrm{P}} = \frac{(3\sqrt{\pi}e)}{2\pi \textrm{c}\omega_{\textrm{LO}}\textrm{m}^*\alpha} \frac{\textrm{sinh}{(\beta/2)}}{\beta^{5/2}}\frac{w^3}{v^3}\frac{1}{K}
\label{eq5}
\end{split}
\end{equation}
where, $\beta$ = hc$\omega_{\textrm{LO}}$/$k_{\textrm{B}}$T, e is the electronic charge, m$^*$ is the effective mass of charge carrier, $\textit{w}$ and $\textit{v}$ correspond to temperature dependent variational parameters. $\textit{K}$ is a function of $\textit{v}$, $\textit{w}$, and $\beta$~\cite{hellwarth1999mobility} i.e., defined as follows: 
 \begin{equation}
\begin{split}
K(a,b) = \int_0^\infty du[u^2 + a^2 - b\textrm{cos}(vu)]^{-3/2}\textrm{cos}(u)
\label{eq6}
\end{split}
\end{equation}
Here, a$^2$ and b are calculated as:
\begin{equation}
\begin{split}
a^2 = (\beta /2)^2 + \frac{(v^2 - w^2)}{w^2v} \beta coth(\beta v/2)
\label{eq7}
\end{split}
\end{equation} 
\begin{equation}
\begin{split}
b =  \frac{(v^2 - w^2)}{w^2v}\frac{\beta}{\textrm{sinh}(\beta v/2)}
\label{eq8}
\end{split}
\end{equation} 

We have used lowest frequency of LO phonon i.e,  1.34, 1.46 and 1.36 THz for Ba$_2$ZrS$_4$, Ba$_3$Zr$_2$S$_7$ and Ba$_4$Zr$_3$S$_{10}$, respectively for our calculation. $\mu_{\textrm{P}}$ gives the upper limit of the charge carrier mobility, under the assumption that charge carrier (electron) interact only with the optical phonon. A significant change in the mobility of charge carrier can be seen on comparing the mobility of electron without including its interaction with the optical phonon (see Table~\ref{Table2}) and with including interaction with the optical phonon (see Table~\ref{Table3}).  Here in Table \ref{Table3}, ionic screening is indicative of the ionicity for a system. On comparing our results with the hybrid inorganic-organic halide perovskites (ionic screening of MAPbI$_3$,  MAPbBr$_3$ and MAPbCl$_3$ are 0.17, 0.18 and 0.22, respectively~\cite{sendner2016optical}), we can say that Ba$_2$ZrS$_4$,  Ba$_3$Zr$_2$S$_7$ and Ba$_4$Zr$_3$S$_{10}$ are less ionic than MAPbX$_3$ (X = Cl, Br, I).  Also, the obtained coupling constant is comparable or larger than MAPbX$_3$ (see Table \ref{Table3}). The lowering of mobility of charge carriers on the inclusion of LO phonon modes indicate that optical phonon modes are dominating over the acoustical phonon modes in these materials. Note that, in the absence of experimental data, these results may help as guideline for further research. Moreover, for qualitative analysis, our results are very informative to understand the charge transport properties of these RP phases. From Table\ref{Table3}, we can clearly see that on increasing n in Ba$_{\textrm{n+1}}$Zr$_\textrm{n}$S$_{\textrm{3n+1}}$ (n=[1-3]) i.e., down the column the polaron mobility decreases and for bulk BaZrS$_3$~\cite{majumdar2020emerging} phase it is very small.  In view of this, the considered RP phases are expected to be better optical material than their bulk phase. 

In conclusion, we have reported the electronic and excitonic properties of the RP phases of Ba$_{\textrm{n+1}}$Zr$_\textrm{n}$S$_{\textrm{3n+1}}$ (n=[1-3]) using Many Body Perturbation Theory. The exciton binding energy decreases on increasing the thickness of the perovskite layer. Double peak character is observed in the first excitonic peak calculated in in-plane direction of Ba$_2$ZrS$_4$. Only Ba$_2$ZrS$_4$ possesses direct band gap and with increasing n in Ba$_{\textrm{n+1}}$Zr$_\textrm{n}$S$_{\textrm{3n+1}}$ (n=[1-3]) the band gap becomes more indirect.  Using Wannier-Mott approach, we have obtained the upper and lower bound of E$_\textrm{B}$, from the electronic and ionic contribution of the dielectric constant, respectively.
We have observed that ionic contribution is more significant in Ba$_{\textrm{n+1}}$Zr$_\textrm{n}$S$_{\textrm{3n+1}}$ than in bulk BaZrS$_3$.
The charge carrier mobility is maximum in Ba$_2$ZrS$_4$, as computed by employing deformation potential of the same.  Further, amongst Ba$_{\textrm{n+1}}$Zr$_\textrm{n}$S$_{\textrm{3n+1}}$ and bulk BaZrS$_3$, the electron-phonon coupling constant is relatively smaller for former RP phases. From our polaron study, we conclude that the optical phonon modes are dominating as compared to the acoustical phonon modes for these systems. A large discrepancy is noticed in the mobility of charge carriers (which includes the effect of acoustical phonon modes only in electron-phonon coupling) and polaron mobility (which includes the effect of optical phonon modes in addition to the acoustic modes in electron-phonon coupling). It shows the dominating character of optical phonon modes in the electron-phonon coupling and must be studied to understand charge transport properties of RP phases. 

\section{Computational Methods}
We have executed a systematic study to explore the optical, electronic and excitonic properties using Density Functional Theory (DFT)~\cite{kohn1965self,hohenberg1964inhomogeneous} and beyond approaches under the framework of Many Body Perturbation Theory~\cite{jiang2012electronic, fuchs2008efficient, basera2019self}.  All calculations are performed with Projected Augmented Wave (PAW) potentials as implemented in Vienna $\textit{Ab initio}$ Simulation Package (VASP)~\cite{kresse1996efficiency,kresse1999ultrasoft}. The PAW potential of elements  viz., Ba, Zr and S contain ten, twelve and six valence electrons, respectively. Ba$_{\textrm{n+1}}$Zr$_{\textrm{n}}$S$_{\textrm{3n+1}}$ (n=[1-3]) RP phases are tetragonal structure having space group I4/mmm [139]. All the structures are optimized using Generalized Gradient Approximation (GGA) as implemented in PBE~\cite{perdew1996generalized}  exchange-correlation ($\epsilon_{\textrm{xc}}$) functional until the forces are smaller than 0.001 eV/\AA. The $\Gamma$-centered 2$\times$2$\times$2 k-mesh sampling is employed for optimization calculations (optimized structures are shown in Fig. \ref{1}). The electronic self-consistency loop convergence is set to 0.01 meV, and the kinetic energy cutoff is set to 600 eV for plane wave basis set expansion. To explore the optical properties and excitonic effects,  Bethe-Salpeter Equation (BSE) is solved. Initially, we have used light 4$\times$4$\times$1 k-mesh for energy calculation (see Fig. S1). The convergence criteria for the number of occupied and unoccupied bands in BSE calculations is given in SI (see Fig. S2). In order to have improved spectral features with denser k-mesh, we have employed the model-BSE (mBSE)~\cite{bokdam2016role} approach. Following this,  we have performed Density Functional Perturbation Theory (DFPT)~\cite{gajdovs2006linear}  with k-mesh 12$\times$12$\times$1, to discern the role of ionic contribution to dielectric function along with electronic contribution.  Note that for GW and BSE calculations, we have used converged NBANDS i.e., 800. Lastly, by employing Frohlich model approach~\cite{frost2017calculating}, we have studied polaron effect in our systems
\begin{acknowledgement}
DG acknowledges UGC, India, for the senior research fellowship [grant no. [1268/(CSIR-UGC NET JUNE 2018)]]. AS acknowledges IIT Delhi for the financial support. MJ acknowledges CSIR, India, for the senior research fellowship [grant no. [09/086(1344)/2018-EMR-I]]. SB acknowledges the financial support from SERB under core research grant (grant no. CRG/2019/000647). We acknowledge the High Performance Computing (HPC) facility at IIT Delhi for computational resources.
\end{acknowledgement}
\begin{suppinfo}
K-mesh convergence for PBE functional; Number of occupied (NO) and unoccupied (NV) bands convergence in BSE calculation; Band structure plot with PBE and PBE+SOC; BSE@G$_0$W$_0$@HSE06 for Ba$_2$ZrS$_4$; model-BSE (mBSE) approach; Projected density of states (PDOS); Effective mass of electron (e) and hole (h); Calculation of deformation potential energy and elastic modulus; Strength of electron-phonon coupling.
\end{suppinfo}

\begin{center}
{\Large \bf Supplemental Material}\\ 
\end{center}
\begin{enumerate}[\bf I.]
	
	\item k-mesh convergence for PBE functional
	\item Number of occupied (NO) and unoccupied (NV) bands convergence in BSE calculation
	\item Band structure plot with PBE and PBE+SOC
	\item BSE@G$_0$W$_0$@HSE06 for Ba$_2$ZrS$_4$
	\item model-BSE (mBSE) approach
	\item Projected density of states (PDOS) 
	\item Effective mass of electron (e) and hole (h)
	\item Calculation of deformation potential energy and elastic modulus
	\item Strength of electron-phonon coupling

\end{enumerate}
\vspace*{12pt}
\clearpage
\section{k-mesh convergence for PBE functional}

\begin{figure}[h!]
 	\centering
 	\includegraphics[width=1.0\textwidth]{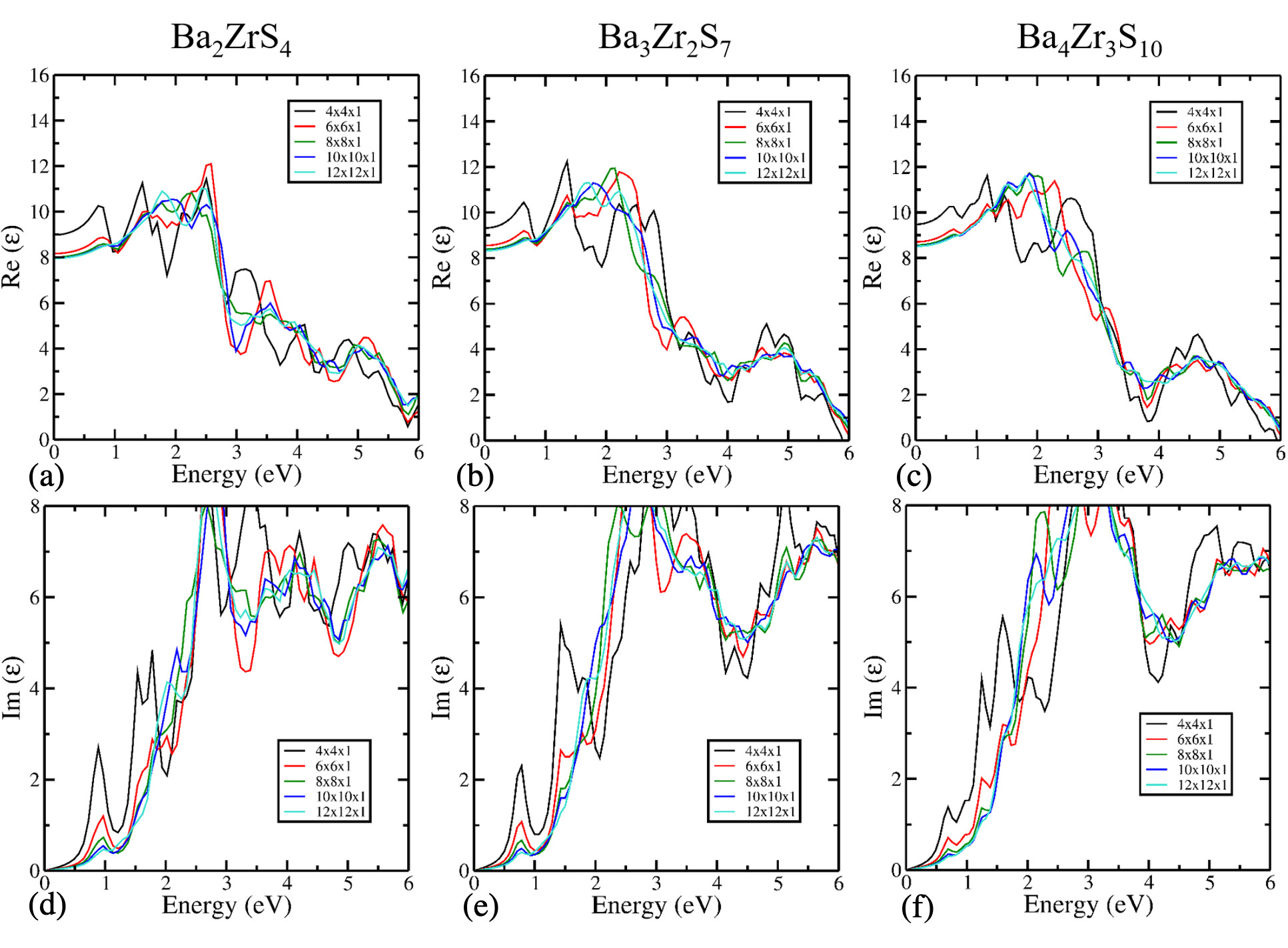}
 	\caption{Real (Re($\varepsilon$)) and imaginary (Im($\varepsilon$)) part of dielectric function for Ba$_{\textrm{n+1}}$Zr$_\textrm{n}$S$_{\textrm{3n+1}}$ (n=[1-3]) RP phases using PBE exchange-correlation $\epsilon_{xc}$ functional.}
 	\label{1}
 \end{figure}
Figure \ref{1}(a-c) shows the variation of real part (Re($\varepsilon$)) and Figure \ref{1}(d-f) shows the imaginary part (Im($\varepsilon$)) (see Figure \ref{1}(d-e)) of dielectric function for  Ba$_{\textrm{n+1}}$Zr$_\textrm{n}$S$_{\textrm{3n+1}}$ (n=[1-3]) RP phases.  On increasing k-mesh a significant change in Re($\varepsilon$) can be seen. However,  no shift in first peak position of Im($\varepsilon$) part of dielectric function is observed on increasing k-mesh. Hence,  4x4x1 light k-mesh is sufficient to compute quasi particle band gap. 

 \section{Number of occupied (NO) and unoccupied (NV) bands convergence in BSE calculation}

\begin{figure}[h!]
	\centering
	\includegraphics[width=0.4\textwidth]{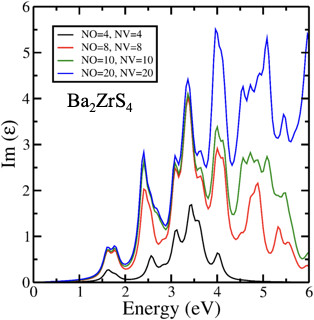}
	\caption{Variation of imaginary part (Im($\varepsilon$)) of dielectric function with number of occupied (NO) and unoccupied (NV) bands using BSE for Ba$_2$ZrS$_4$.}
	\label{2}
\end{figure}
\newpage
\section{Band structure plot with PBE and PBE+SOC}

\begin{figure}[h!]
	\centering
	\includegraphics[width=1.0\textwidth]{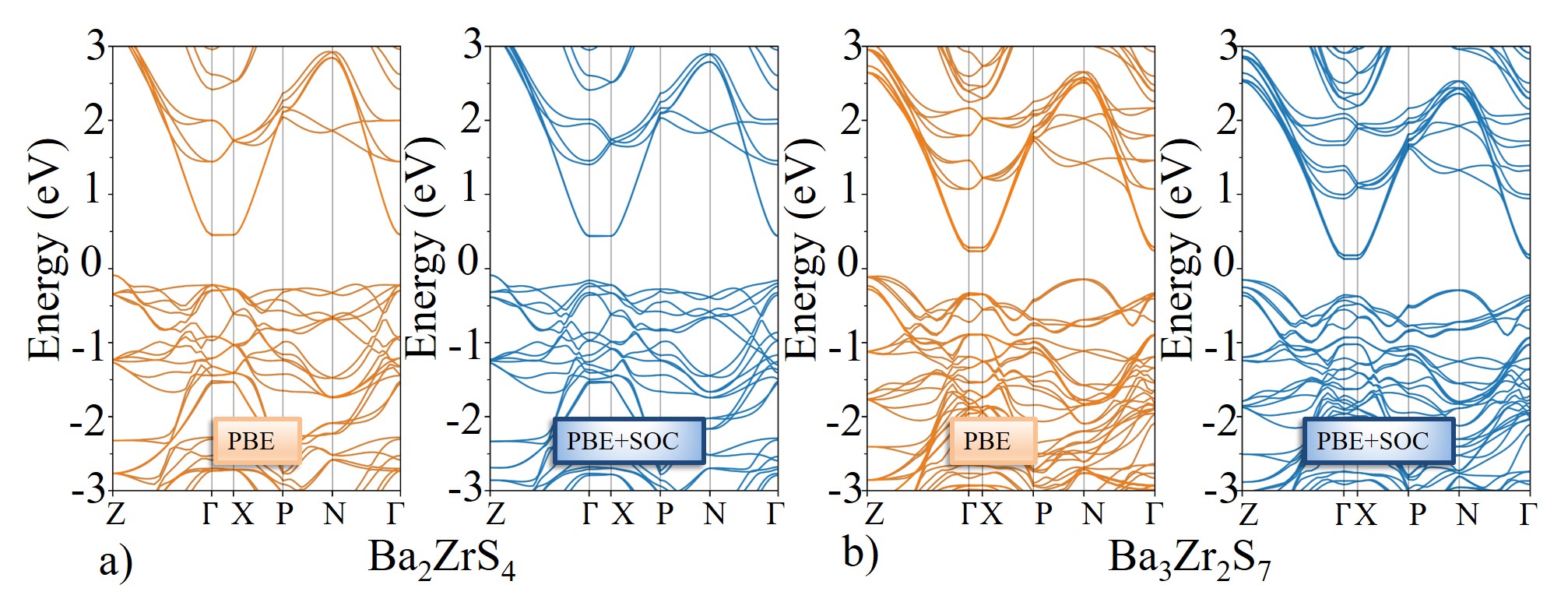}
	\caption{Band structure of (a) Ba$_2$ZrS$_4$, (b) Ba$_3$Zr$_2$S$_7$ using PBE and PBE+SOC exchange-correlation $\epsilon_{\textrm{xc}}$ functional.}
	\label{3}
\end{figure}

\newpage
\section{BSE@G$_0$W$_0$@HSE06 for Ba$_2$ZrS$_4$}

\begin{figure}[h!]
	\centering
	\includegraphics[width=0.6\textwidth]{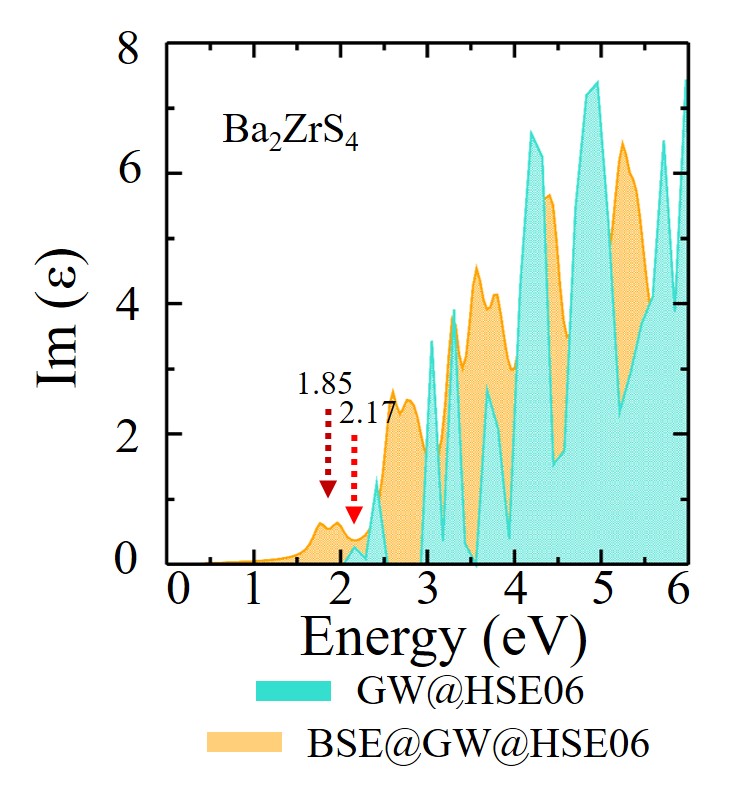}
	\caption{Imaginary part (Im ($\epsilon$)) of the dielectric function for Ba$_2$ZrS$_4$ using G$_0$W$_0$@HSE06 and BSE@G$_0$W$_0$@HSE06.}
	\label{4}
\end{figure}

\newpage
\section{model-BSE (mBSE) approach}
\begin{figure}[h!]
	\centering
	\includegraphics[width=0.60\textwidth]{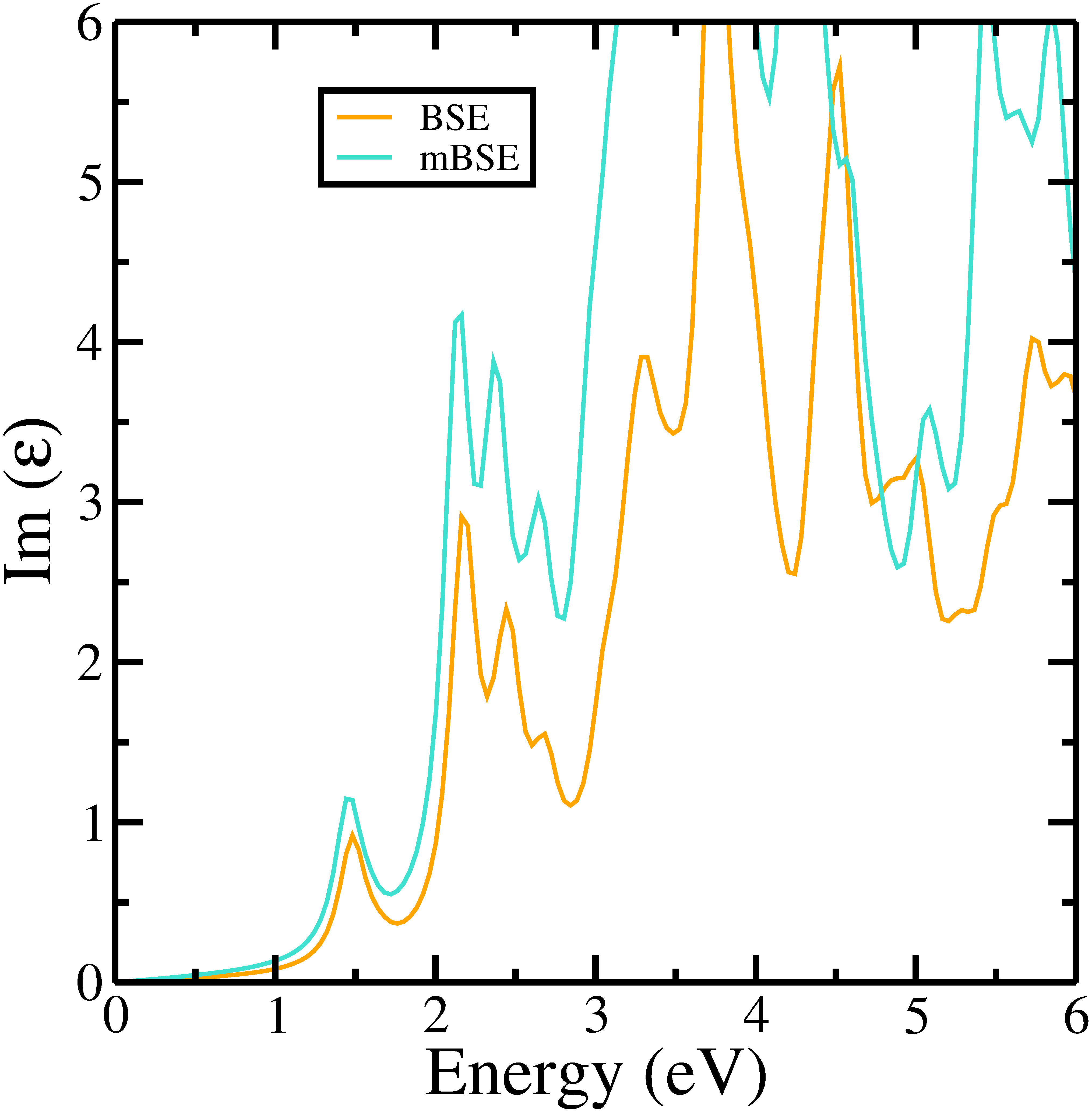}
	\caption{Imaginary part (Im ($\varepsilon$)) of the dielectric functional for Ba$_3$Zr$_2$S$_7$ using BSE and mBSE.}
	\label{6}
\end{figure}

To compute exciton energy and E$_\textrm{B}$ precisely, one needs to accurately calculate the optical spectra or optical gap using conventional BSE@G$_0$W$_0$ approach. However, there is inconsistency observed in BSE exciton peak, due to insufficient high k-mesh (approx 8$\times$8$\times$1). We can not use high k-mesh because it is computationally very expensive. This results in the incorrect E$_\textrm{B}$ value. Therefore, to overcome this issue, a less expensive but robust mBSE approach was proposed. In this model, the convergence of the optical spectra as a function of the number of k-mesh is performed. This method is generally based on two approximations:

(i) Using Eq.~\ref{eq1} the RPA static screening W is replaced by a simple analytical model. Here, the dielectric function is replaced by the local model function:
\begin{equation}
\begin{split}
\varepsilon^{-1}_{\textrm{G,G}}(\textrm{$|\textrm{q+G}|$})=1-(1-\varepsilon^{-1}_{\infty})\exp(-\dfrac{\textrm{$|\textrm{q+G}|$}^2}{4\lambda^2})
\label{eq1}
\end{split}
\end{equation}\\
where, $\varepsilon_\infty$ is the static ion-clamped dielectric function in the high frequency limit. $\varepsilon_{\infty}^{-1}$ is calculated either from DFPT or G$_0$W$_0$. q and G are the wave vector and lattice vector of the reciprocal cell, respectively. $\lambda$ is the screening length parameter, calculated by fitting $\varepsilon^{-1}$ at small wave vectors with respect to $|\textrm{q+G}|$ (see Figure~\ref{6} (a-c)). The parameters obtained for the  Ba$_{\textrm{n+1}}$Zr$_\textrm{n}$S$_{\textrm{3n+1}}$ (n=[1-3]) RP phases are collected in Table~\ref{Table1}.
\newpage
\begin{table}[htbp]
	\caption{The calculated inverse of static ion-clamped dielectric function $\varepsilon_{\infty}^{-1}$ and the screening length parameter $\lambda$ (\AA$^{-1}$) used in mBSE (Eq. \ref{eq1}) for Ba$_{\textrm{n+1}}$Zr$_\textrm{n}$S$_{\textrm{3n+1}}$ (n=[1-3]) RP phases.} 
	\begin{center}
		\begin{tabular}[c]{|c|c|c|} \hline
			 Ba$_{\textrm{n+1}}$Zr$_\textrm{n}$S$_{\textrm{3n+1}}$ & $\varepsilon_{\infty}^{-1}$ (PBE) & $\lambda$ (PBE) \\ \hline
			Ba$_2$ZrS$_4$ & 0.14 & 1.17 \\ \hline
			Ba$_3$Zr$_2$S$_7$ & 0.15 & 1.19  \\ \hline
			Ba$_4$Zr$_3$S$_{10}$ & 0.15& 1.17  \\ \hline
		\end{tabular}
		\label{Table1}
	\end{center}
\end{table}
\begin{figure}[h!]
	\centering
	\includegraphics[width=1.0\textwidth]{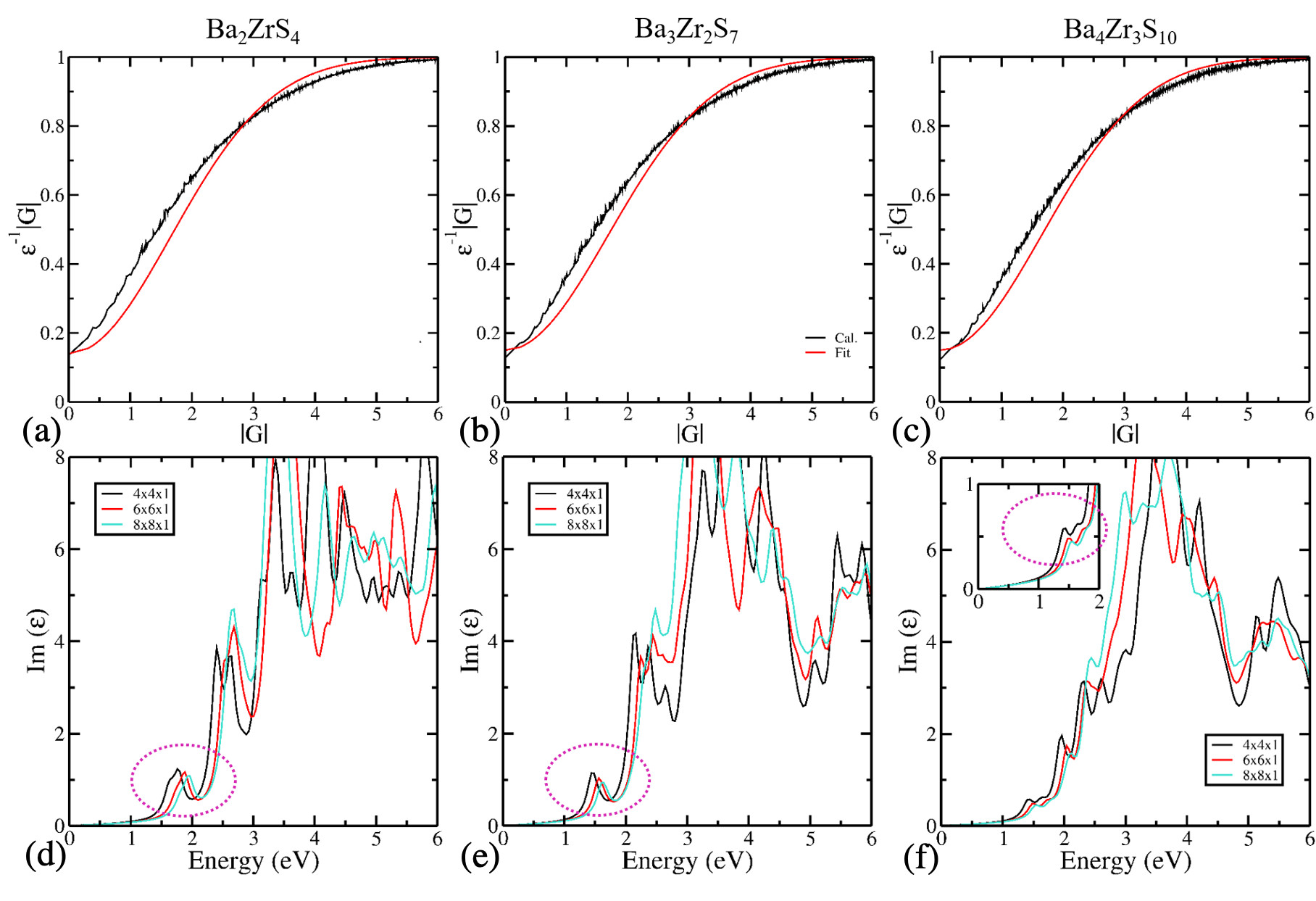}
	\caption{Variation of inverse of the dielectric function $\varepsilon^{-1}$ with respect to $|\textrm{q+G}|$ for (a) Ba$_2$ZrS$_4$, (b) Ba$_3$Zr$_2$S$_7$, and (c) Ba$_4$Zr$_3$S$_{10}$, respectively. The red curve is obtained by fitting based on Eq. (5). The mBSE calculated spectra with different k-mesh for (d) Ba$_2$ZrS$_4$, (e) Ba$_3$Zr$_2$S$_7$ and (f) Ba$_4$Zr$_3$S$_{10}$, respectively.}
	\label{5}
\end{figure}

Here, in Figure ~\ref{6}, we have shown that the imaginary part of the dielectric function calculated with BSE@GW@PBE, matches with the one, which is calculated with the model BSE (mBSE) method, where we have chosen PBE as the starting point. Therefore, we can say that the excitonic features (i.e first peak) information are always retained by mBSE approach. Notably, both the calculations are performed using 4$\times$4$\times$1 k-mesh with the same starting point.

\newpage
\section{Projected density of states (PDOS)}
\begin{figure}[h!]
	\centering
	\includegraphics[width=1.0\textwidth]{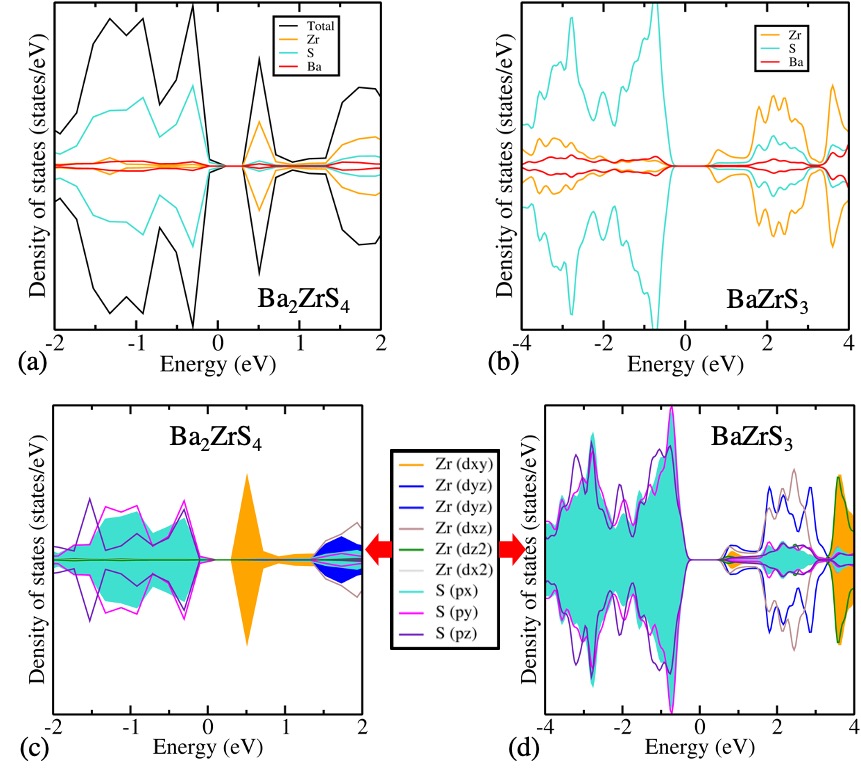}
	\caption{Partial density of states (pDOS) for (a) Ba$_2$ZrS$_4$ and (b) BaZrS$_3$. Orbital contribution of S and Zr in VBM and CBm, respectively for (c) Ba$_2$ZrS$_4$ and (d) BaZrS$_3$. }
	\label{7}
\end{figure}
\newpage
\section{Effective mass of electron (e) and hole (h)}
\begin{figure}[h!]
	\centering
	\includegraphics[width=1.0\textwidth]{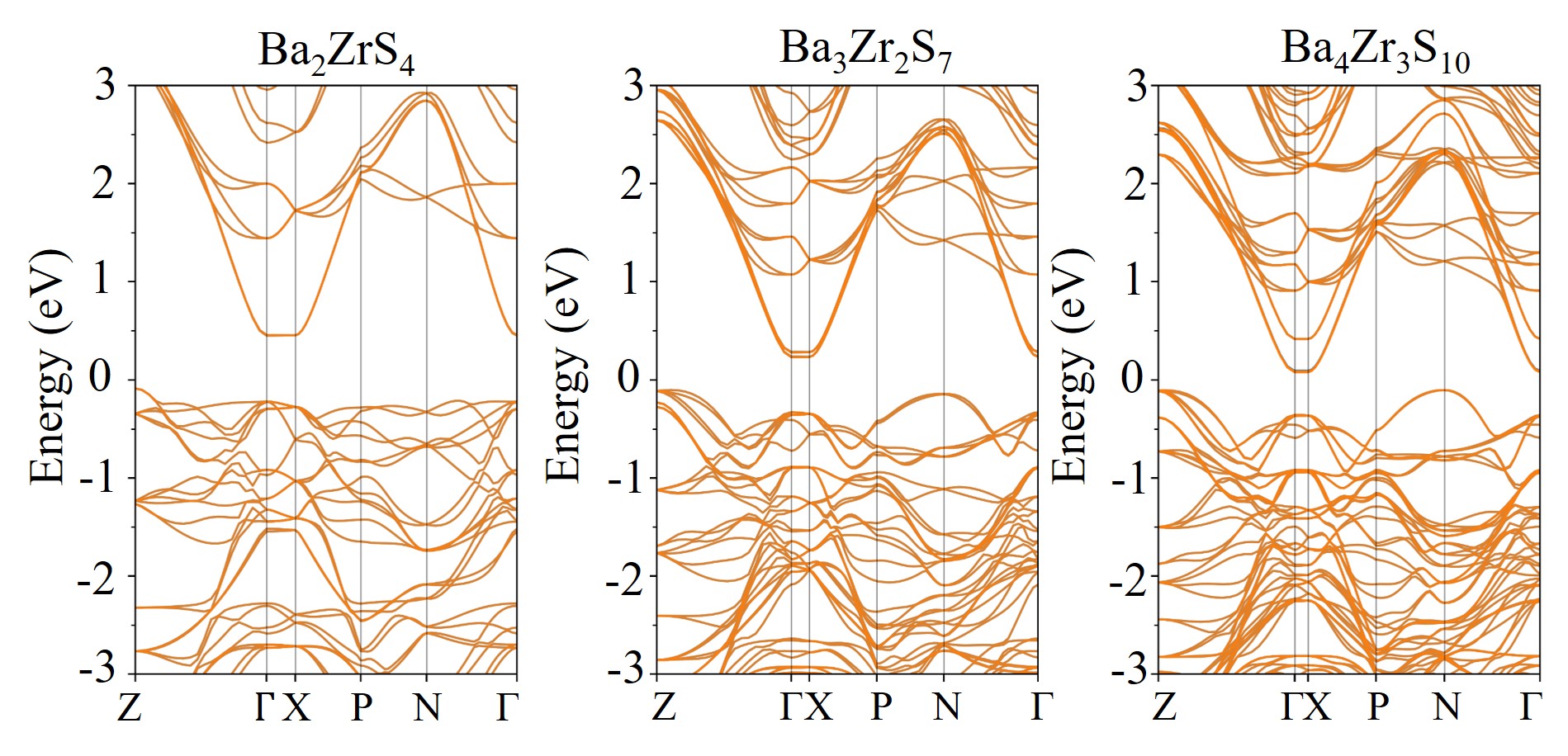}
	\caption{Band structure of  Ba$_{\textrm{n+1}}$Zr$_\textrm{n}$S$_{\textrm{3n+1}}$ (n=[1-3]).}
	\label{8}
\end{figure}
To compute effective mass, we have obtained the bandstructure using PBE $\epsilon_{xc}$ functional (see Figure \ref{10}). Here, we have calculated the effective masses of electron (e) and hole (h) along the symmetric path $\Gamma$ $\rightarrow$ Z for Ba$_2$ZrS$_4$.  In case of Ba$_3$Z$_2$S$_7$  and Ba$_4$Zr$_3$S$_{10}$, the effective masses of e and h are calculated along $\Gamma$ $\rightarrow$ Z and Z $\rightarrow$ $\Gamma$, respectively. All the effective and reduced masses are tabulated in Table \ref{Table2}.  Note that,  in Figure \ref{10}, the observed flatbands along high-symmetric path $\Gamma$ $\rightarrow$ X are along z-direction. While calculating the effective masses, we have excluded the high-symmetric path along z-direction.

\begin{table}[htbp]
	\caption{Effective masses of electron, hole and reduced mass (in term of rest mass of electron (m$_\textrm{e}$)) for Ba$_{\textrm{n+1}}$Zr$_\textrm{n}$S$_{\textrm{3n+1}}$ (n=[1-3]) RP phases.} 
	\begin{center}
		\begin{tabular}[c]{|c|c|c|c|} \hline
			 Ba$_{\textrm{n+1}}$Zr$_\textrm{n}$S$_{\textrm{3n+1}}$ & m$_{\textrm{e}}^{*}$ & m$_{\textrm{h}}^{*}$ & $\mu$ \\ \hline
			Ba$_2$ZrS$_4$ & 0.18 & 0.43& 0.32 \\ \hline
			Ba$_3$Zr$_2$S$_7$ & 0.23 & 1.85 & 0.26  \\ \hline
			Ba$_4$Zr$_3$S$_{10}$ & 0.26 & 1.25 & 0.34  \\ \hline
		\end{tabular}
		\label{Table2}
	\end{center}
\end{table}
\newpage
\section{Calculation of deformation potential energy and elastic modulus}
\begin{figure}[h!]
	\centering
	\includegraphics[width=1.0\textwidth]{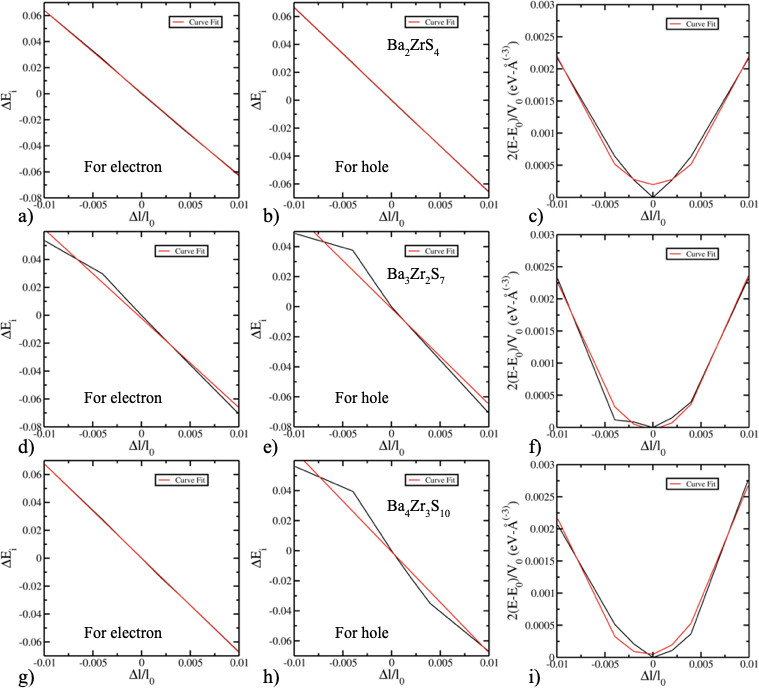}
	\caption{Deformation potential and elastic modulus of RP phases.}
	\label{10}
\end{figure}

According to deformation potential model the carrier mobility is defined as: 
\begin{equation}
\begin{split}
\mu_{\textrm{DP}} = \frac{(8\pi)^{\frac{1}{2}}\hbar^4e\textrm{C}_{\textrm{3D}}}{3(m^*)^{5/2}(k_\textrm{B}\textrm{T})^{3/2}\textrm{E}_{1}^2}
\label{eq2}
\end{split}
\end{equation}
where $k_{\textrm{B}}$,  T,  and m$^*$ correspond to the Boltzmann constant, temperature (i.e., 300 K), and effective mass, respectively. In the above equation, E$_1$ corresponds to the deformation potential constant along y-direction for electron and hole, respectively.  It is given by:
\begin{equation}
\begin{split}
\textrm{E}_1 = \frac{\Delta \textrm{E}_\textrm{i}}{\Delta \textrm{l}/\textrm{l}_0}
\label{eq3}
\end{split}
\end{equation}
where, $\Delta \textrm{E}_\textrm{i}$ denotes the change in the energy or shift in the position of VBM or CBm under uniaxial strain along the y-direction. l$_0$ is the lattice constant along the transport direction. $\Delta \textrm{l}$ denotes the change or deformation in the lattice constant l$_0$ on application of uniaxial starin.  The elastic modulus C$_{3\textrm{D}}$ is computed using (E - E$_0$)/V$_0$ = C($\Delta \textrm{l}/\textrm{l}_0)^2/2$, where E$_0$ and E are the total energy of undeformed system and deformed system. V$_0$ represents the equilibrium volume of the system. Note, herein, we have chosen strain range from -1.0\% to +1.0\% to obtain the fitted values of C$_{3\textrm{D}}$ and E$_1$ (see Figure \ref{10}).

\begin{figure}[h!]
	\centering
	\includegraphics[width=0.60\textwidth]{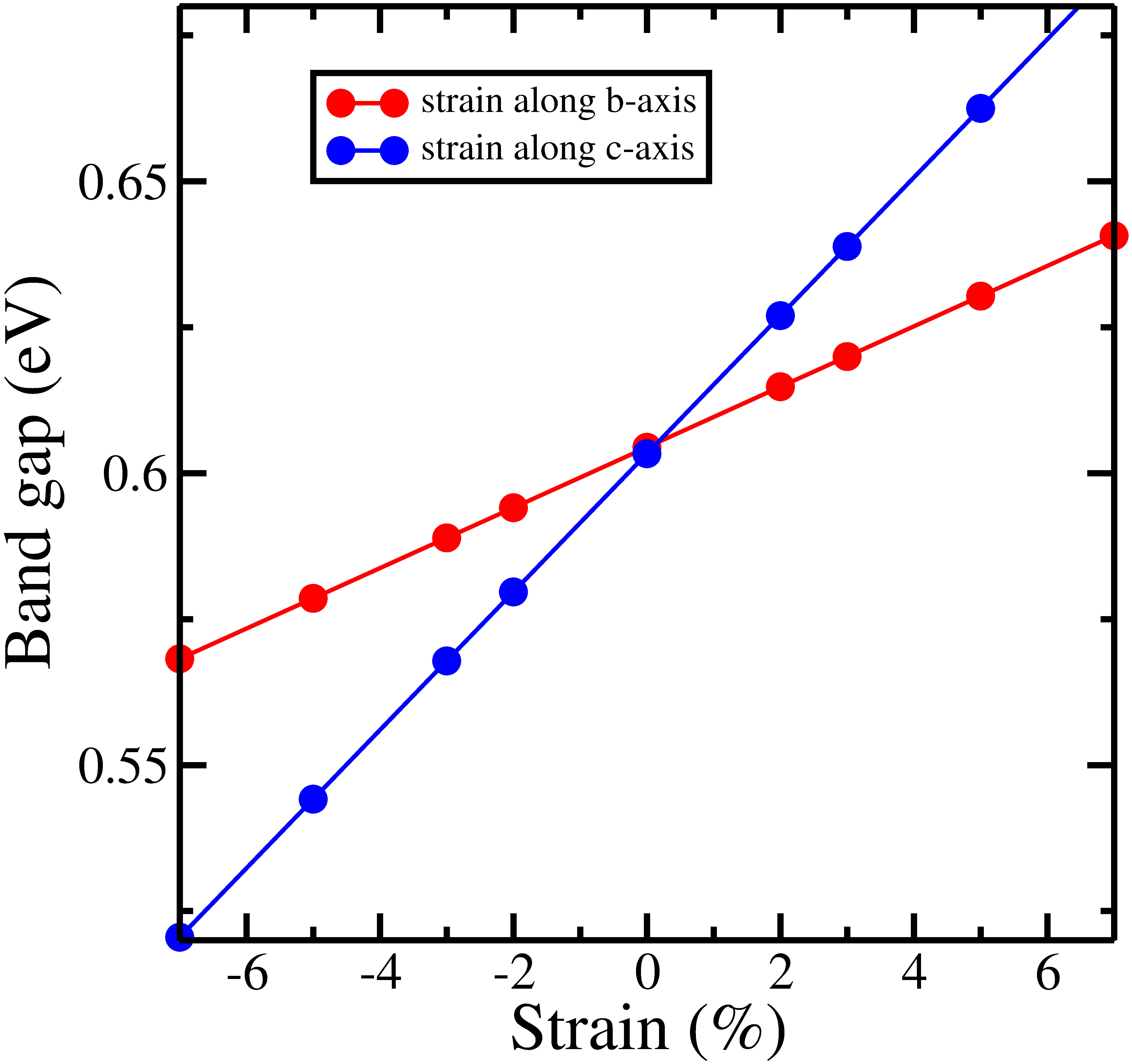}
	\caption{Band gap variation with strain along b- and c-direction for Ba$_2$ZrS$_4$ using PBE $\epsilon_{\textrm{xc}}$.}
	\label{9}
\end{figure}

\newpage
\section{Strength of electron-phonon coupling}
For the qualitative idea of electron-phonon coupling strength, we have calculated the specific free volume.
 It is already reported in the literature that electron-phonon coupling depends on free volume (unoccupied space)~\cite{he2018unravelling}.  Lattice free volume is defined as the difference between the unitcell volume and the constituent ions' volume. Following this, the ratio between the unoccupied volume and the total volume is defined as the specific free volume (see Table ~\ref{Table3}).  From Table \ref{Table3}, we can clearly see that specific free volume is quite large and hence the electron-phonon coupling could be prompt in these systems.  Hence, electron-phonon coupling is important in these systems.
\begin{table}[h!]
	\caption{Specific free volume of Ba$_{\textrm{n+1}}$Zr$_\textrm{n}$S$_{\textrm{3n+1}}$ (n=[1-3]).} 
	\begin{center}
		\begin{tabular}[c]{|c|c|c|c|} \hline
				Ba$_{\textrm{n+1}}$Zr$_\textrm{n}$S$_{\textrm{3n+1}}$ & Specific free volume (\%)\\ \hline
			Ba$_2$ZrS$_4$ &  94.33  \\ \hline
			Ba$_3$Zr$_2$S$_7$ &  94.17  \\ \hline
			Ba$_4$Zr$_3$S$_{10}$ &  94.10 \\ \hline
		\end{tabular}
		\label{Table3}
	\end{center}
\end{table}
\bibliography{achemso-demo}
\end{document}